# Enhanced Support Vector Machine Based Signal Recovery in Bandwidth-Limited 50-100 Gbit/s Flexible DS-PON

**Liyan Wu[1], Yanlu Huang[1], Kai Jin[1], Shangya Han[1], Kun Xu[1], Yanni Ou[1]***

[1]*State Key Laboratory of Information Photonics and Optical Communications, Beijing University of Posts and Telecommunications, Beijing 100876, China, *yanni.ou@bupt.edu.cn*

**Abstract:** We proposed an adaptive signal recovery algorithm with reduced complexity based on the SVM principle for flexible downstream PON. Experimental results indicate a record-high link power budget of 24 dB for bandwidth-limited 100 Gbit/s direct-detection transmission@1E-3.

## 1. Introduction

To meet the future high-speed network demands of 5/6G technologies, cost-effective high-speed transmission systems with a rate of 100 Gbit/s or beyond are required. Considering the OPex/CaPex and the passive optical networks (PON) evolution legacies, developing high-speed PON utilizing lower bandwidth devices is one of the viable solutions [1,2]. Accordingly, the flexible data rate with different modulation formats and code-rate selections for PONs was proposed in [1], enabling 50 Gbit/s nonreturn-to-zero (NRZ) and 100 Gbit/s 4-level pulse amplitude modulation (PAM-4) bandwidth-limited intensity modulation and direct-detection (IM/DD) transmission in downstream PON (DS-PON). Inevitably, bandwidth-limited induced inter-symbol interference (ISI) combined with dispersion will necessitate the use of digital signal processing (DSP) algorithms for effective signal recovery. Given that the optical network units (ONUs) are sensitive to DSP complexity, designing DSP approaches that can effectively recover the distorted signals in high-speed systems while also satisfying the complexity constraints of DS-PON is becoming increasingly important.

Previous research has focused on signal recovery on both the transmitter and receiver sides of IM/DD systems. Meanwhile, according to [1], reducing the achievable forward error correction (FEC) threshold from 1.9E-2 to 1.8E-3 can increase the net bit rate by 15.56 Gbit/s. Therefore, a lower FEC threshold is desirable when evaluating the performance of signal recovery algorithms. In [3] and [4], link power budgets of 22 dB and 17 dB were achieved at the FEC threshold of 1E-3, by employing the transmitter side constant-modulus-algorithm (CMA) and the receiver side Volterra nonlinear equalization (VNE). In [5], a nonlinear Tomlinson-Harashima precoding and a VNE of 83 taps were also proposed, achieving -18 dB receiver sensitivity at the FEC threshold of 1E-3. However, these algorithms suffer from relatively high complexities which are hard to be implemented in ONUs. Meanwhile, algorithms using relatively lower complexity feedforward equalizer (FFE) and feedback equalizer (DFE) algorithms, e.g., 47-tap FFE and 25-tap DFE configurations, find difficulties in achieving either a better FEC threshold of 1E-3 [5] or a high link power budget [6-7] when recovering the received signals. To address the challenges, the equalization based on support vector machine (SVM) algorithms were proposed in [8,9] as SVM-I and SVM-II, which employ different feature vector (FV) composition structures superior to FFE&DFE. However, the algorithms were demonstrated so far only in the fixed 50 Gbit/s PON, while their modifications and evaluations to fit into a higher speed PON system with the flexible rate and bandwidth-limited channel characteristics have not yet been performed.

In this work, by modifying the conventional SVM-I and SVM-II principle, an enhanced SVM signal recovery algorithm with reduced complexity is proposed to fit into the ONU receiver side, in order to recover the bandwidth-limitation and dispersion impaired high-speed flexible-rate signals in a feasible manner. The proposed algorithm is then evaluated experimentally on a DS-PON testbed to analyze its recovery performances for both 50 Gbit/s NRZ and 100 Gbit/s PAM-4 transmissions over 40 km fibers using 25 G-class optoelectronic devices, emulating a bandwidth-limited high-speed flexible-rate PON scenario. By applying the proposed enhanced SVM algorithm solely at the receiver side without any transmitter pre-emphasis, the ONU receiver sensitivity is improved by 2.3 dB and 3.3 dB compared to the conventional SVM-I/II algorithms and FFE/DFE algorithms, respectively. Moreover, experimental results demonstrate an improvement of over 2 dB in the link power budget compared to the highest value reported in [1] and [3-7]. To the best of our knowledge, this achieves a record-high link power budget of 24 dB for the bandwidth-limited 100 Gbit/s direct-detection transmission at a FEC threshold as low as 1E-3, utilizing 25G-class optoelectronic devices with a semiconductor optical amplifier (SOA) adopted solely at the receiver side.

## 2. Principle of the proposed SVM algorithm

The principle of the SVM algorithm consists of the training process to obtain the optimal hyperplane as shown in Fig.1(a), and the test process for the received signals decisioning based on the trained hyperplane as shown in Fig. 1(b). In the training process, the transmitted samples $\overrightarrow{Tx}$ before the DAC and the detected samples $\vec{X}$ after the ADC, as shown in Fig. 1(c), are both fed into the training label block and the training feature vector (FV) block respectively, which are treated as inputs of the SVM solver block to output the normal vector $\vec{w}$ and the bias constant $b$, forming the classification hyperplane $\vec{w^T} \cdot \vec{X} + b$. Once the training process is completed, the following samples will be fed into the FV block in the test process, where the FV is input to the Hyperplane for signal decisioning. Fig. 1(g) shows the hyperplane configured for deciding NRZ and PAM4. To form an enhanced SVM algorithm (SVM-Proposed), a new FV composition approach is proposed as well as the new SVM solver configuration. Fig. 1(d) compares the FV composition of the SVM-Proposed method with that of the conventional SVM-I and SVM-II methods, where the proposed FV considers both the feedforward and feedback samples ISI. Based on the proposed FV scheme, the Training FV and the Test FV are composed as shown in Fig. 1(e), where $x(k)$ is the received $k^{th}$ symbol, $t_x(k)$ is the corresponding $k^{th}$ training label for the original symbol, $d(k)$ is the detected result and is fed back into the test FV for further convergence. To reduce the complexity, the adaptive configurations of SVM solver algorithms are proposed to recover the signals in different modulation formats respectively, considering a flexible PON scenario. Fig. 1(f) illustrates the Stochastic Gradient Descent (SGD) algorithm chosen for recovering NRZ, and the Sequential Minimal Optimization (SMO) algorithm for recovering PAM4. Both SGD and SMO provide the reduced complexity equals to or slightly higher than $O(N)$ of FFE/DFE [10], while significantly lower than $O(N + N^2)$ of VNE shown in Fig. 1(h).

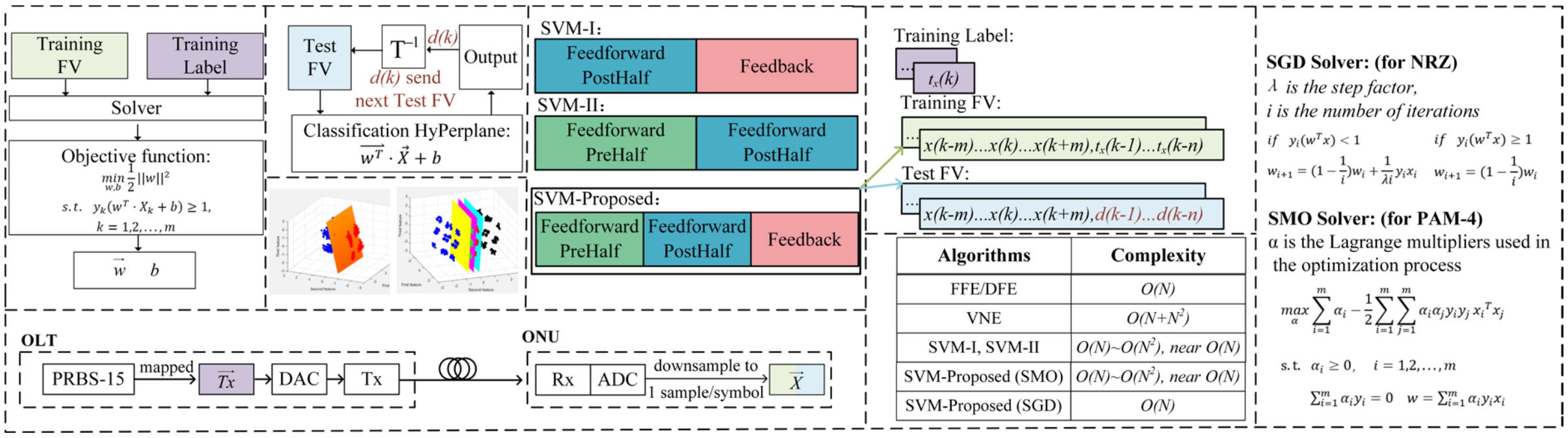

Fig. 1. The proposed SVM algorithm principle, (a) training process to obtain the optimal hyperplane parameters, (b) test process for signals decisioning, (c) training and test sets $\vec{Tx}$ and $\vec{X}$ obtained from the testbed, (d) different FV compositions, (e) details of training FVs, labels, and test FVs, (f) SMO and SGD solvers for different SVM algorithms, (g) NRZ and PAM-4 after recovery, (h) algorithms complexity comparison.

## 3. Experimental setup

To evaluate the recovery performance of the proposed enhanced SVM algorithm, a direct-detection flexible DS-PON experimental testbed is established to support the bandwidth-limited transmission of the 50 Gbit/s NRZ and 100 Gbit/s PAM-4 based on the 25G-class optoelectronic devices as shown in Fig. 2. The 3dB bandwidth of the overall system frequency response is measured as around 13 GHz as indicated in Fig. 2 (e). The optical line termination (OLT) contains an arbitrary waveform generator (AWG), an electrical amplifier (EA), a distributed feedback laser at 1310.5 nm, and a 25G-class intensity modulator (IM). A 1:2 splitter is to connect ONU1 and ONU2 located at different fiber distances of F-I and F-II, which can both be set between 0/40 km. Each ONU contains two variable optical attenuators (VOA-I and II), an SOA, a commercial photodetector (PD) integrated with a trans-impedance amplifier (TIA), and a real-time Oscilloscope (OSC). VOA-I is to emulate the varying received optical power, while the VOA-II is set fixed to prevent the PD from being struck. The offline DSP at the OLT included symbol generation and root-raised-cosine (RRC) filter, while at the ONU side, time recovery and equalization are included. NRZ and PAM-4 signals have different tap settings. For NRZ, both FFE/DFE and SVM-Proposed (SMO, SGD) use 9 feedforward and 3 feedback taps, while for PAM-4 31 feedforward and 5 feedback taps were utilized to improve the equalization ability.

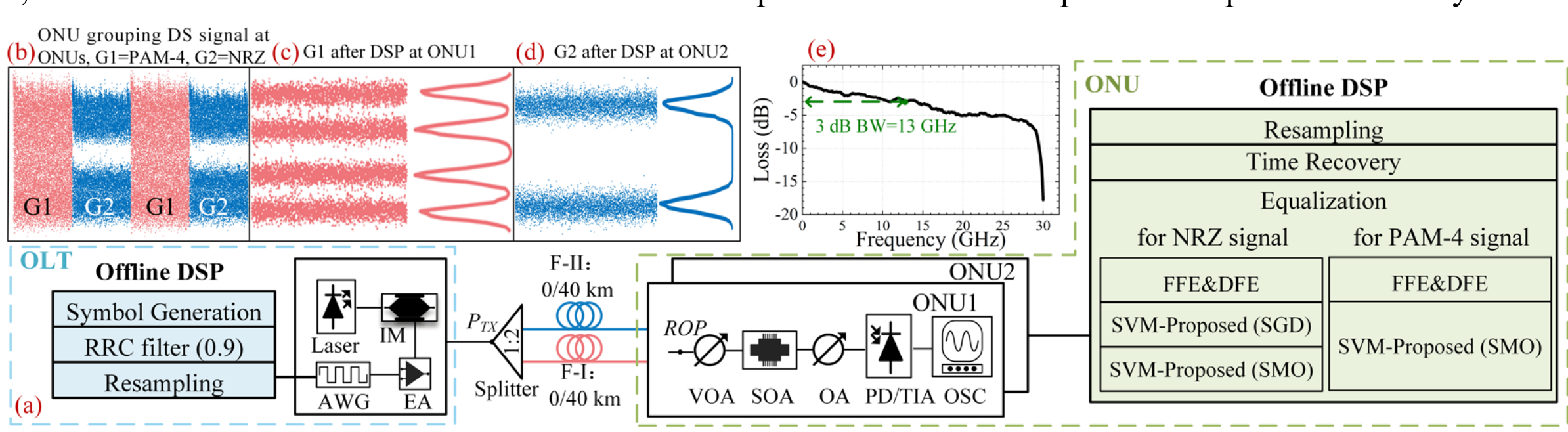

Fig. 2. Experimental setup, (a) testbed composition and the offline DSP details at both OLT and ONUs, including 92 GSa/s 32GHz AWG, 25-G-class IM, 1310 nm laser, 20G-PD with TIA, and 80 GSa/s 30 GHz real-time OSC, (b) downstream signals received at one of the ONUs after clock recovery but before equalization, (c, d) signals received by ONU1 and 2 after equalization, (e) channel frequency response of OLT-ONU1.

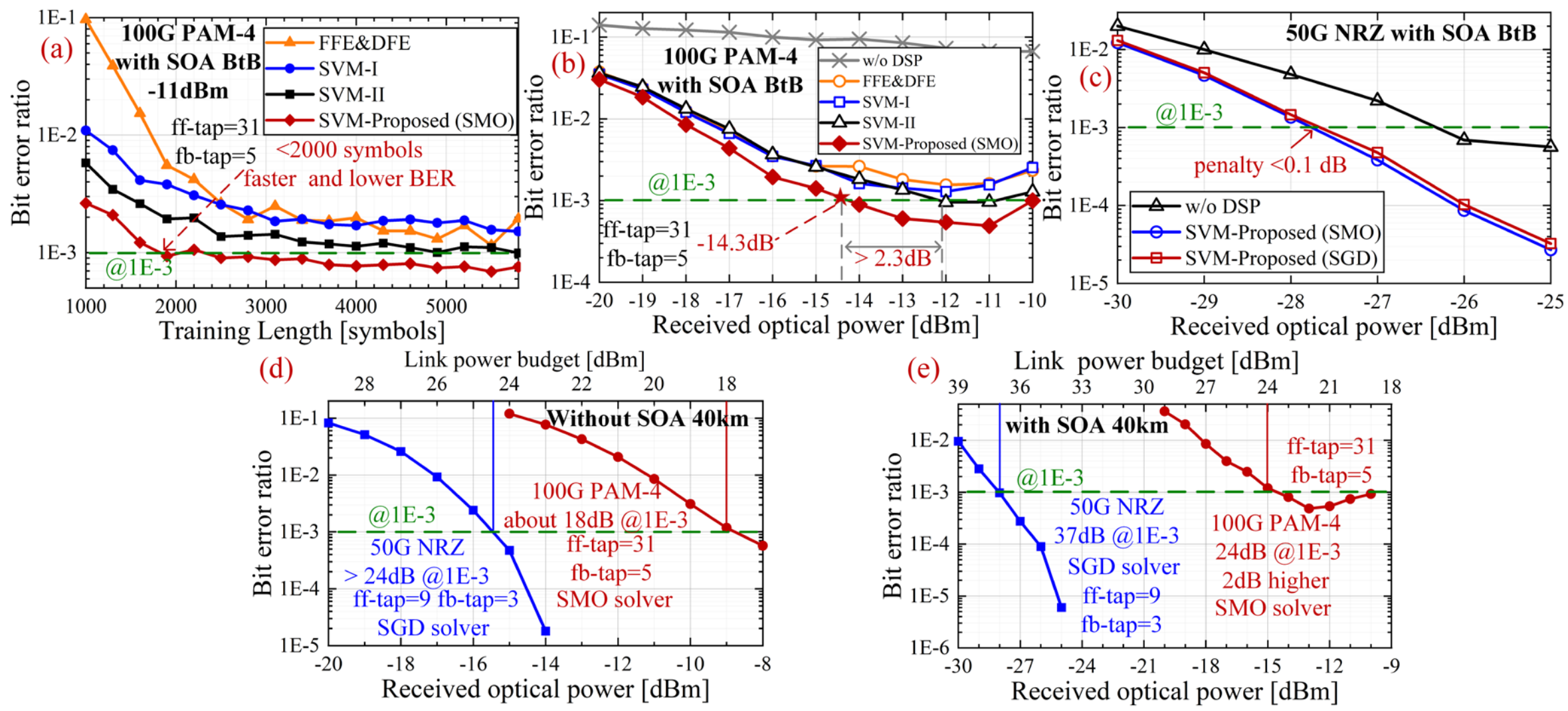

Fig. 3. Comparisons of (a) taps convergence time among algorithms, (b) receiver sensitivities among algorithms, improvement >2.3 dB, (c) different SGD solvers for NRZ detection, penalty <0.1dB, (d-e) link power budgets @ FEC of 1E-3, without and with the SOA.

## 4. Results

To evaluate the feasibility of the SVM-Proposed algorithm, equalization taps convergence time, receiver sensitivity, and link power budget of the IM/DD are analyzed, as well as the feasibility of SVM solvers for different modulation formats. Fig. 3(a) firstly analyzed the tap convergence time, where 31 feedforward taps and 5 feedback taps are used. Results indicate that the SVM-Proposed algorithm requires fewer than half of the convergence iterations (from ~5000 to ~2000 symbols), achieving faster tap convergence time and more optimized converged values (BER < 1E-3).

Fig. 3(b) evaluates the receiver sensitivity when using the 31+5 taps length across all the algorithms for received signals recovery. The SVM-Proposed (SMO) algorithm achieves an ROP of -14.3 dBm, showing an improvement of 2.3 dB over SVM-I/-II and an enhancement of 3.3 dB over FFE/DFE. This performance improvement is attributed to the algorithm's accurate channel ISI characterization. Then, Fig. 3(c) analyzes the ROP versus BER performance of the proposed SVM with reduced complexity in the 50 Gbit/s NRZ transmission. Compared to the SVM-Proposed (SMO) algorithm, the SVM-Proposed (SGD) algorithm achieves a complexity reduction of $O(N)$, which is equivalent to that of FFE/DFE. This approach delivers a performance comparable to the SVM-Proposed (SMO) algorithm while incurring an ROP penalty of less than 0.1dB only. Furthermore, in Fig. 3(d) and (e), the link power budgets over a 40 km transmission with and without the SOA at the ONU receiver side are analyzed respectively, the results show that, without the SOA, the link power budgets for 50 Gbit/s NRZ using the SVM-Proposed (SGD) algorithm and for 100 Gbit/s PAM-4 using the SVM-Proposed (SMO) algorithm can achieve 24 dB and 18 dB, respectively. Meanwhile, with the SOA solely at the receiver side, a 24 dB link power budget for 100 Gbit/s PAM-4 is achieved. To the best of our knowledge, this achieves a record-high link power budget for the bandwidth-limited 100 Gbit/s direct-detection transmission at a FEC threshold as low as 1E-3 under the same system condition, indicating an over 2 dB improvement compared to the highest value reported in [1] and [3-7]. Meanwhile, for NRZ, the link power budget over 37 dB is achieved, indicating that both the 50 Gbit/s and 100 Gbit/s transmission can be adequately supported using the proposed SVM algorithms without introducing additional complexities or performance degradations.

## 5. Conclusions

We proposed an enhanced SVM algorithm with reduced complexity of $O(N)$ to address the bandwidth-limited induced impairments on the 50-100 Gbit/s flexible DS-PON. The performance of the proposed algorithm was experimentally evaluated in terms of equalization taps convergence time, receiver sensitivity, link power budget, computational complexity, and SVM solvers feasibility, etc. The SVM-Proposed algorithm can achieve faster convergence times, and receiver sensitivity improvements of 2.3 dB and 3.3 dB compared to the conventional SVM-I/-II and FFE/DFE algorithms. Moreover, a record-high link power budget of 24 dB was achieved for the 100 Gbit/s bandwidth-limited IM/DD DS-PON transmission with an SOA adopted solely at the receiver side.